\documentclass[pop,aip,twocolumn,floatfix,amsmath,amssymb,showpacs,tightenlines]{revtex4}
\usepackage{mathrsfs}
\usepackage{bbm}
\usepackage{amssymb}
\usepackage{graphicx,color}
\usepackage{epsfig,graphicx,times}
\usepackage{amstext}
\usepackage{amsmath}
\usepackage{latexsym}
\usepackage{bm}
\usepackage{CJK}

\begin{document}

\title{Foam  Au driven by  4$\omega$ - 2$\omega$  ignition laser pulse  for inertial confinement fusion}
\author{Ke Lan \email{lan_ke@iapcm.ac.cn}}
\author{Peng Song}

\affiliation{Institute of Applied Physics and Computational
Mathematics, Beijing, 100094, China}

\begin{abstract}
Green light (2$\omega$) has the potential to drive ignition target
for laser fusion with significantly more energy than blue light
(3$\omega$) and a relatively higher damage threshold for the optic
components in the final optic assembly, but it has
 issues of a relatively low laser to x-ray conversion
efficiency and a hard x-ray spectrum as compared to 3$\omega$. In
this paper, we propose to drive a foam  hohlraum wall  with an
ignition laser pulse by taking a 4$\omega$ laser
  at the pre-pulse and a 2$\omega$ laser at the main-pulse, called as 4$\omega$ - 2$\omega$  ignition  pulse.
This novel design has the following advantages: (1) benefiting from
2$\omega$ of its relatively high energy output and low damage
threshold  during main-pulse; (2) benefiting from foam   in its
relatively high laser to x-ray conversion efficiency and  relatively
low M-band fraction in re-emission; (3)  benefiting from 4$\omega$
of its low LPI during pre-pulse. From our 1D simulations with Au
material, the laser to x-ray conversion in a foam driven by
4$\omega$ - 2$\omega$ pulse  has an increase of $28\%$ as compared
to a solid target driven by 3$\omega$ with the same   pulse shape.
The relatively thin optical depth of foam is one of the main reasons
for the increase of laser to x-ray conversion efficiency inside a
foam target.

\end{abstract}


\maketitle

\section{Introduction}

The choice of laser wavelength is central to the  inertial
confinement fusion (ICF) indirect-drive ignition study, because
laser plasma instabilities (LPI) impacts a wide range of physics
issues, such as hohlraum energetics, implosion symmetry and pulse
shaping\cite{lindl2003}. Present laser facilities
 designed for ICF study almost all operate at blue light
(3$\omega$), such as the National Ignition Facility\cite{lindl2014},
the Shenguang series \cite{zhengwanguo, lanke2016}, the
OAMEG\cite{Campbell} and the Laser M$\acute{e}$gajoule facility
\cite{LMJ}, because 3$\omega$ has the advantages of a higher laser
deposition\cite{Neumayer, Kline2011} and a lower LPI
\cite{Froula2007} than green light (2$\omega$). However, 3$\omega$
is generated via the frequency-tripling technique with a relatively
low efficiency, and in addition, it has a relatively low damage
threshold for the optic components in the final optic assembly
\cite{Haynam}. The two issues greatly limit the maximum energy
output capability of a 3$\omega$ laser facility. In contrast,
2$\omega$ laser can highly increase the threshold for optical
damages, which therefore  results in a lower operation costs for a
laser facility and a higher laser energy deliverable on target
\cite{Suter, Niemann}. Furthermore, in the long term, any reactor
based on laser-driven ICF will require laser operation at
sufficiently high wavelength for keeping the live time of the optics
at a level that meets the energy production requirements
\cite{Depierreux}. In fact, a series experimental works with
2$\omega$ laser beams \cite{Stevenson, Labaune, Depierreux2012,
Lizhichao} have been performed, and they demonstrated that LPI
aroused by 2$\omega$ is in an acceptable level for fusion
experiments. Thus, 2$\omega$ may hold significant advantages for
future ignition and high-fusion yield study and can be a laser
wavelength competitive with  3$\omega$ for future ignition laser
facility. Nevertheless, 2$\omega$ laser has a lower laser to X-ray
conversion efficiency  than 3$\omega$ due to its lower laser
deposition and  higher LPI backscatters, which therefore requires
higher laser energy and  power. These issues seriously block the use
of 2$\omega$ for an ignition laser facility.

In indirect-drive ICF study, the energy coupling from laser to
capsule can be described by three efficiency: the absorbed laser
efficiency $\eta_{aL}$, the laser to x-ray conversion efficiency
$\eta_{LX}$,  and the hohlraum to capsule $\eta_{HC}$. Here,
$\eta_{aL}$  is decided by the  laser deposition via inverse
bremsstrahlung process and the backscatters due to LPI, $\eta_{LX}$
is decided by atomic processes, bremsstrahlung emission and its
inverse process, and $\eta_{HC}$  is decided by hohlraum geometrical
parameters and albedos of hohlraum wall and capsule. Notice that
$\eta_{aL}$ and $\eta_{LX}$ are related to both laser  and   target
parameters, while $\eta_{HC}$ is only connected  to the target
parameters including wall materials and geometrical configurations
 at a given radiation environment inside a hohlraum. At a
longer laser wavelength $\lambda_L$, both $\eta_{aL}$ and
$\eta_{LX}$ are lower, which is due to   lower laser deposition and
  higher LPI for $\eta_{aL}$ while  due to  thinner laser
deposition depth for $\eta_{LX}$. As a result, for a given target,
2$\omega$ laser has a lower $\eta_{aL}$ and a lower $\eta_{LX}$ than
3$\omega$ and 4$\omega$. However,  both $\eta_{aL}$ and $\eta_{LX}$
are also connected to the target parameters, such the initial
density $\rho_0$, one can therefore consider to change $\rho_0$ to
increase $\eta_{aL}$ and $\eta_{LX}$ at 2$\omega$ to make them be
comparable with those at 3$\omega$ with a solid target.

In fact, it was predicted by HR analytical theory that the lower
density hohlraum walls produce higher radiation temperature than the
high density walls \cite{HR2005} because it can reduce hydrodynamic
losses. According to HR theory, the radiation front propagates
subsonically in materials with high density and part of the absorbed
energy is wasted by the flow kinetic energy, while it propagates
supersonic in the materials with lower density and can devote almost
all of the absorbed energy to heating the material. From this
theory, a foam target has a higher albedo, and it therefore has  a
higher $\eta_{HC}$ than a solid target. This prediction was
demonstrated successfully by later experiments with Ta$_2$O$_5$
foams\cite{HR2008} and  Au foams\cite{Dongyunsong2013, Zhanglu2015,
Zhanglu2016} and numerical simulations with Au
foams\cite{Zhanglu2011, Dongyunsong2013-1}. Notice that above
experimental and numerical comparisons of foam and solid were
performed under the same lasers at 3$\omega$. However, to explore
the possibilities of   2$\omega$ laser for future ignition laser
facilities, we need to know: how about the x-ray output of a foam Au
target under a 2$\omega$ laser as compared to a solid Au target
under  3$\omega$? It would be  encouraging  for  2$\omega$ laser if
its x-ray output  can be higher, or at least not lower, than
3$\omega$. Otherwise, it   would be discouraging for  2$\omega$
laser   in ICF application. In addition,
 the foams used in above  comparisons have   thicknesses covering
both laser ablation region and radiation ablation region, which
combines the influences of  foam on $\eta_{aL}$, $\eta_{LX}$, and
$\eta_{HC}$. We need to make clear if a foam can help  to increase
$\eta_{aL}$ and $\eta_{LX}$ at 2$\omega$ as compared to those of a
solid target at 3$\omega$ or not.

In this paper, based on our one-dimensional simulations   of x-ray
output from Au targets with different $\rho_0$ driven by an ignition
laser pulse with various $\lambda_L$, we  propose to drive a foam
hohlraum wall with an ignition laser pulse at 4$\omega$ during its
pre-pulse and 2$\omega$  during   main-pulse for in-direct drive
ICF, called as 4$\omega$-2$\omega$ pulse hereafter. Here, it is
worth to mention that 4$\omega$ laser has the highest $\eta_{aL}$
and $\eta_{LX}$ with  the lowest LPI than 2$\omega$ and 3$\omega$
for the same target, but meanwhile, it has the lowest damage
threshold for the optic components. To utmost utilize the advantages
of 4$\omega$ laser to greatly inhibit the hydrodynamic instabilities
aroused by LPI during pre-pulse, we therefore consider to use
4$\omega$ laser for the pre-pulse. Moreover, to  make clear whether
a foam can help to increase $\eta_{aL}$ and $\eta_{LX}$ at 2$\omega$
as compared to the case of a solid target at 3$\omega$ or not, we
will use a foam-solid target, in which the foam part faces to laser
source and   has a thickness approximately equal to   the laser
ablated depth under the drive laser. The foam-solid target has the
same areal density as the solid target.

The remaining presentation is organized as follows. In Sec. II, we
will present the code and models used in this study. In Sec. III, we
will discuss the simulation results of different models by comparing
their   laser to X-ray conversion efficiency and the M-band
fraction, and we analyze the reasons why foam can help to increase
the laser to x-ray conversion efficiency. Finally, we will present a
summary in Sec. IV.

\section{CODE and MODEL}

We  use our one-dimensional (1D) multi-group radiation hydrodynamic
code RDMG \cite{Feng1999} to simulate the Au plane targets, foam or
solid, under an ignition laser pulse. RDMG is widely used in both
indirect-drive  and direct-drive ICF studies\cite{Xuyan2006,
Huo2012, Xuyan}, including both theoretical
  and experimental studies on Shenguang series laser facility. In
RDMG, we solve the two-temperature hydrodynamic equations coupled
with a multi-group radiation transfer equation. The multi-group
radiation transfer equation is solved with S-N discrete coordinate
scheme, and the energy coupling of radiation with matter is resolved
by the matrix operator splitting method \cite{Feng1998}. The laser
energy deposition via inverse bremsstrahlung is calculated with a
three-dimensional ray tracing package. The electron thermal
conduction is treated by the Spitzer-H$\ddot{a}$rm model
\cite{Cohen, Spitzer}with a flux limiter \cite{Malone} which is
usually taken as 0.08 in our simulations. The thermodynamic
quantities are derived either from the ideal gas model or from data
of realistic equation of state. In this work, we take 120 groups in
solving the multi-group radiation transfer equation.

\begin{table}[htbp]
\caption{Parameters of laser and target in all models.  Here, the
laser pulse of  $4\omega$-$2\omega$ means that it takes $4\omega$
laser for the pre-pulse and $2\omega$ laser for the main-pulse. The
foam-solid Au is composed by two parts: a foam part with $\rho_0$ =
0.05 g/cm$^3$ and $\Delta$ = 40 $\mu$m, and a solid part with
$\rho_0$ = 19.24 g/cm$^3$ and $\Delta$ = 39 $\mu$m, with the foam
part facing to the laser source. For the solid Au, $\rho_0$ = 19.24
g/cm$^3$ and $\Delta$ = 40 $\mu$m. }
\begin{tabular}{cccc}
  \hline
  \hline
  Model & Laser pulse & Au plane target \\
  \hline
  I &   $3\omega$ & Solid  \\
  II &   $4\omega$ & Solid   \\
  III &   $2\omega$ & Solid   \\
   IV & $4\omega$-$2\omega$  & foam-solid\\
  \hline
  \hline
\end{tabular}
\label{expdata}
\end{table}

\begin{figure}[htbp]
\includegraphics[bb=170 150 420 550,width=4 cm]{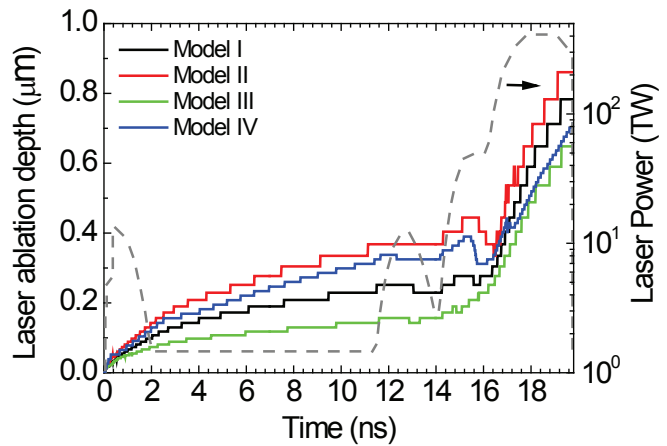}
\caption{(color online) Ignition pulse shape used in CH Rev5 design
\cite{Haan} (dashed grey line) and temporal evolutions of the  laser
ablation depth for all models (solid lines).}
\label{laser depth}
\end{figure}

We consider a typical ignition laser pulse shape used in CH Rev5
design for the National Ignition Facility \cite{Haan}, which is
presented in Fig. \ref{laser depth}. The peak laser intensity is
taken as $5 \times 10^{14}$ W/cm$^2$. As shown, the pre-pulse
finishes at around 16 ns where it is 49 TW, and then the main-pulse
begins. Therefore, our $4\omega$-$2\omega$ pulse for this work is
$4\omega$ laser before 16 ns and  $2\omega$ laser thereafter. The
four models we consider in simulations are given in the table. In
model I, II and III, it is a solid Au plane target driven by
2$\omega$, 3$\omega$ and 4$\omega$ lasers, respectively. In model
IV, it is a foam-solid Au plane target driven by the
4$\omega$-2$\omega$ laser. The solid target has a thickness $\Delta$
of 40 $\mu$m with $\rho_0$ = 19.24 g/cm$^3$. The "foam-solid" target
is composed by two parts: a foam part with $\rho_0$ = 0.05 g/cm$^3$
and $\Delta$ = 40 $\mu$m and a solid part with $\rho_0$ = 19.24
g/cm$^3$ and $\Delta$ = 39 $\mu$m, with the foam part facing to the
laser source. The areal density of the foam part equals to a solid
layer with 19.24 g/cm$^3$ in density and 1 $\mu$m in thickness. The
reason for taking such a thickness for the foam part is because the
laser ablation depth is approximately 1 $\mu$m for  Au driven by the
CH Rev5 ignition laser pulse.  We use such a thin foam layer only
for the laser deposition region just  to investigation whether the
foam can help to increase the laser to x-ray conversion efficiency
under the $4\omega$-$2\omega$ pulse  as compared to the solid Au
under the $3\omega$ pulse. Moreover, we take  $\Delta$ = 39 $\mu$m
for the solid part of the foam-solid Au while  $\Delta$ = 40 $\mu$m
for the  solid Au, just to keep the same areal density for both
 targets, as mentioned in Sec. I.

\section{SIMULATIONS and DISCUSSIONS}

In in-direct drive approach,  the laser absorption efficiency, the
laser to x-ray conversion efficiency, the x-ray output  and its
 spectrum are very important for hohlraum energetics and capsule implosion
 performances, while they are  strongly connected to both   laser
and  target parameters. In this section, we present
 and compare the simulation results from RDMG for
the four models .

\begin{figure}[htbp]
\includegraphics[bb=170 150 420 550,width=4 cm]{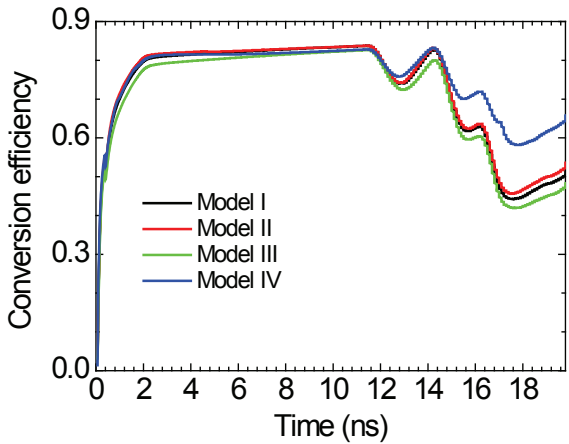}
\caption{(color online) Temporal evolutions of the  laser to x-ray
conversion efficiency for all models.} \label{conversion}
\end{figure}

The laser ablation depth $\triangle_L$ is  decided not only by laser
intensity and laser wavelength, but also by the initial density of
target \cite{MTV}.   Presented in Fig. \ref{laser depth} is
$\triangle_L$ for all models. Here, we use 19.24 g/cm$^3$ as the
nominated density of the foam part and calculate $\triangle_L$ under
this nominated density for model IV.   The results are discussed
below. (1) For the same target, $\triangle_L$ strongly depends on
$\lambda_L$,  and it is smaller at a longer $\lambda_L$. For the
first three models, $\triangle_L$ is about 0.7 $\mu$m at 2$\omega$,
0.78 $\mu$m at 3$\omega$ and 0.86 $\mu$m at 4$\omega$ at the end of
the laser pulse. (2) $\triangle_L$ is more sensitive to $\lambda_L$
during pre-pulse than main-pulse. Fitting from the first three
models, we have $\triangle_L \propto \lambda_L^{-3/2}$ for pre-pulse
and $\triangle_L \propto \lambda_L^{-1/2}$ for the main-pulse. (3)
During the main-pulse, $\triangle_L$ is mainly decided by
$\lambda_L$ while relatively insensitive to  $\rho_0$. Models III
and IV have almost the same $\triangle_L$ because they both use
2$\omega$ during the main-pulse.

\begin{figure}[htbp]
\includegraphics[bb=170 150 420 550,width=4 cm]{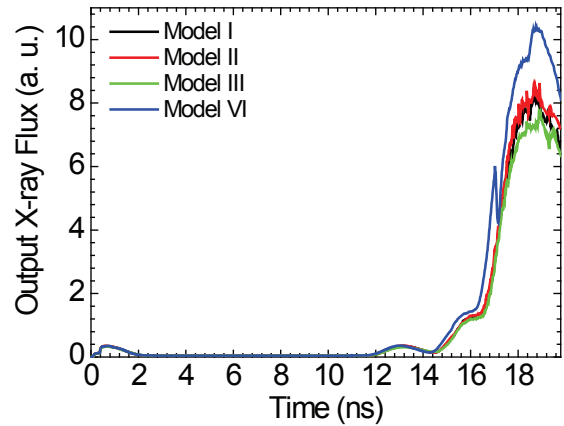}
\caption{(color online) Temporal evolutions of the output x-ray flux
emitted from the target surface where the laser is deposited.}
\label{re-emission_flux}
\end{figure}

\begin{figure}[htbp]
\includegraphics[bb=170 150 420 550,width=4 cm]{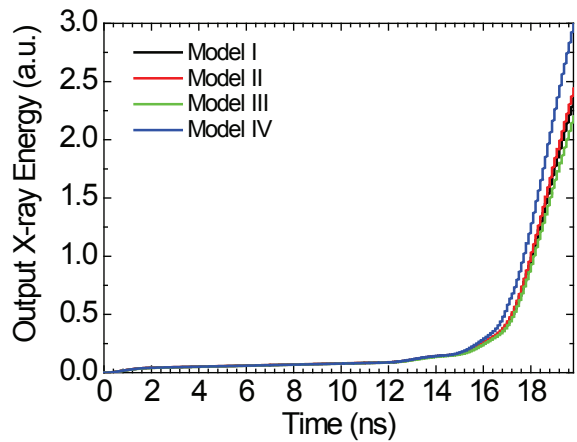}
\caption{(color online) Output x-ray  energy emitted from the target
surface where the laser is deposited.} \label{re-emission_energy}
\end{figure}

For the  laser absorption efficiency $\eta_{aL}$ given in this
paper, we only consider the laser absorption via inverse
bremsstrahlung. The laser backscatter caused by LPI is not taken
into considerations in this work, because present simulation
results, including results from 2D codes, cannot adequately describe
the observed LPI \cite{Kline056314} and therefore a qualitatively
study on backscatter requires the experimental measurements. From
our simulations, $\eta_{aL}$ is higher at a shorter $\lambda_L$, and
it is higher inside the foam-solid Au than inside the solid Au under
the same laser pulse with same $\lambda_L$. Nevertheless,
$\eta_{aL}$ rises sharply as time, reaching $99\%$ in less than 1 ns
and approaching $100\%$ in all models. Therefore, $\eta_{aL}$ has
few difference among the four models.

Presented in Fig. \ref{conversion} is the temporal evolutions of
$\eta_{LX}$ for all models. The results are discussed below. (1)
$\eta_{LX}$ rises rapidly during the first step of the ignition
laser pulse and reaches around $80\%$ at about 2ns for all models.
(2) $\eta_{LX}$ drops at the frontiers of all later steps when the
laser power rises steeply. (3)  $\eta_{LX}$ is lower at a longer
$\lambda_L$, but it can be remarkably increased by the foam layer of
the foam-solid Au. From the results of the first three models which
use solid Au, $\eta_{LX}$ is $52.3\%$ at 4$\omega$, $50\%$ at
3$\omega$ and $47\%$ at 2$\omega$  at the end of the laser pulse.
However, the foam layer of the foam-solid Au can remarkably increase
$\eta_{LX}$ at   2$\omega$, and can make $\eta_{LX}$ even obviously
higher  than 4$\omega$ with solid Au. In model IV, $\eta_{LX}$ is
about $64\%$ at the end of laser pulse, which is $36\%$ higher than
the 2$\omega$ model, $28\%$ higher than the 3$\omega$ model, and
$22\%$ higher than the 4$\omega$ model with solid Au. As a result of
the increased   $\eta_{LX}$ in the foam-solid Au,  both output x-ray
flux and x-ray energy driven by 2$\omega$ laser are be remarkably
increased in model IV, and they can be even higher than the cases
driven by 3$\omega$ and 4$\omega$  with  solid Au. From Fig.
\ref{re-emission_flux} and Fig. \ref{re-emission_energy},
 the output x-ray  flux and x-ray energy of model IV  is about $36\%$
higher than the 2$\omega$ model, $28\%$ higher than the 3$\omega$
model, and  $22\%$ higher than the 4$\omega$ model which all use
solid Au.

\begin{figure}[htbp]
\includegraphics[bb=140 170 390 570,width=4.35 cm]{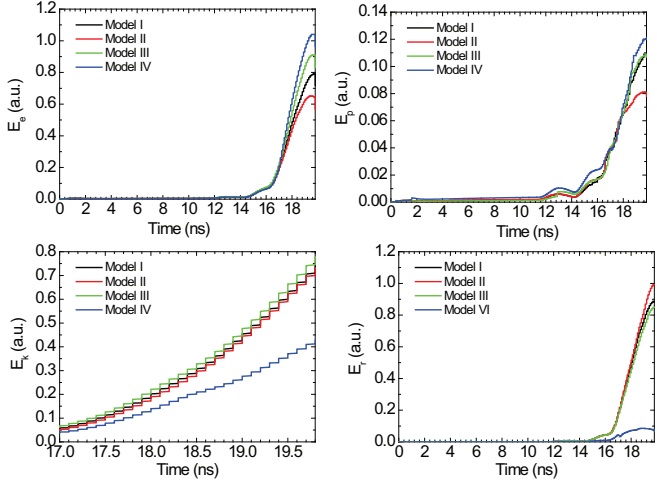}
\caption{(color online) Temporal evolutions of   electronic thermal
energy $E_e$, the ionic potential energy $E_p$, the plasma kinetic
energy $E_k$, and the radiation energy $E_r$ for all models.}
\label{Ee}
\end{figure}

In order to understand why the x-ray output can be increased with a
foam layer, we compare all kinds of energies of the laser ablation
region among the four  models, where the electronic density is lower
than the critical density and  the laser can be deposited. Recall
that in model IV, the foam layer faces to the laser source  and  has
a thickness equal to the laser ablation depth. Hence, the main
influences caused by the foam layer mainly happen in the laser
ablation region. Presented in Fig.\ref{Ee} is temporal evolutions of
the electronic thermal energy $E_e$, the ionic potential energy
$E_p$, the plasma kinetic energy $E_k$, and the radiation energy
$E_r$ for all models. Here, we take an arbitrary unit for these
energies. As shown, all $E_e$, $E_p$, $E_k$ and $E_r$ depend on both
$\lambda_L$ and $\rho_0$, but with very different behaviors. Notice
that  we have neglected the ionic thermal energy in Fig.\ref{Ee}
because it is very small in the laser ablation region and can be
neglected as compared to all other kinds of energies.

According to our simulation results, both $E_e$ and $E_p$ are higher
at a longer $\lambda_L$, and  they are higher in the foam-solid Au
than in the solid Au. From Fig.\ref{Ee},  $E_e$  is 0.92, 0.79 and
0.65 in the solid Au driven by the laser pulses at 2$\omega$,
 3$\omega$ and  4$\omega$, respectively; and it is 1.04 in the foam-solid Au
under the 4$\omega$-2$\omega$ pulse. Again from Fig.\ref{Ee}, $E_p$
is about 0.11, 0.11 and 0.08 in the solid Au driven by the laser
pulses at  2$\omega$,
 3$\omega$ and  4$\omega$, respectively; and it is 0.12 in the foam-solid Au under the
4$\omega$-2$\omega$ pulse. However, the differences of $E_e$ and
$E_p$ among these four models are not remarkable. Nevertheless,
these results are  connected to the electron thermal conduction
model used in our simulations, and it is worth to be checked by
future simulations with more accurate physics models.

\begin{figure}[htbp]
\includegraphics[bb=170 250 480 620,width=8 cm]{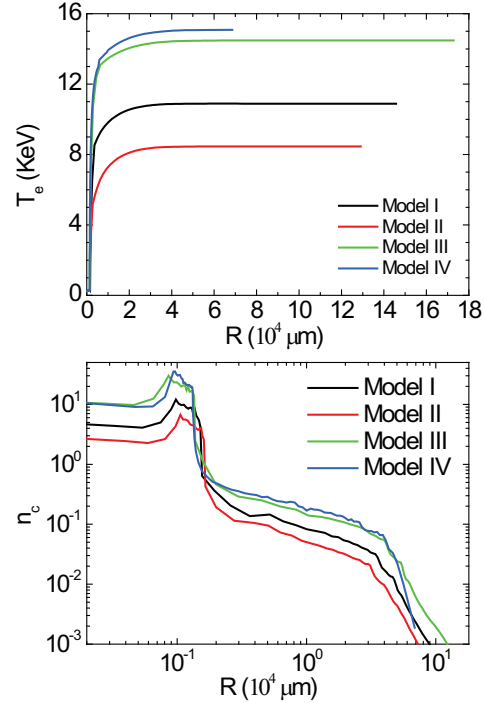}
\caption{(color online) Spatial distributions of electronic
temperature $T_e$ and electronic density $n_e$, which  is normalized
to the critical density of corresponding $\lambda_L$, at 19.5 ns
before the laser pulse ends.} \label{Te and Ne}
\end{figure}

\begin{figure}[htbp]
\includegraphics[bb=170 150 420 550,width=4 cm]{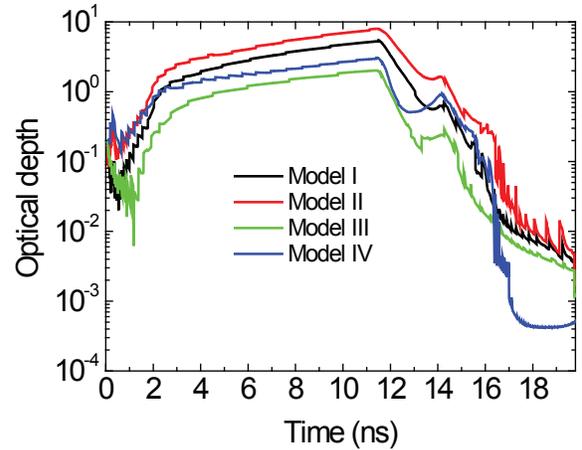}
\caption{(color online) Temporal evolutions of the optical depth for
all models.} \label{optical depth}
\end{figure}

In contrast, both $E_k$ and $E_r$ are remarkably decreased by using
the foam-solid Au as compared to the solid Au. Here, we only compare
model IV with model I which uses solid Au driven by 3$\omega$ laser,
because the differences among the first three models  are small.
From Fig.\ref{Ee},  $E_k$ of model IV is 43$\%$ lower than model I,
and $E_r$ of model IV is about 92$\%$ lower than model I. It means
that in the laser ablation region, the kinetic energy of model IV is
down by about half from model I, and the radiation energy of model
IV decreases by  an order of magnitude over model I.  In the
following, we discuss the reasons for such remarkable decreases of
$E_k$ and $E_r$, respectively.

There are two reasons for  the decrease of $E_k$  in the foam-solid
Au. One reason is due to a thinner  $\Delta_L$ in the foam than in
the solid, as shown in Fig.\ref{laser depth}. Another one is due to
the propagation competition between the radiation front and sonic
speed. According to Refs.\cite{HR2005} and \cite{HR2008}, the
radiation front propagates subsonically and part of the absorbed
energy is wasted by the flow kinetic energy for the solid target,
while the front velocity is supersonic and can devote almost all of
the absorbed energy to heating the material for the foam target.
Presented in Fig.\ref{Te and Ne} is the spatial distributions of
electron temperature $T_e$ and normalized electron density $n_e$
 at 19.5 ns before the laser pulse ends. As presented, model IV with
foam-solid Au has a much shorter expansion than the solid target
models. In addition, $T_e$ is higher at a longer $\lambda_L$; and
$T_e$ in model IV  is close to model III, because they both use
$2\omega$ laser during the main-pulse.  Moreover, the spacial
distributions of $n_e$ are sensitive to  $\lambda_L$, and  $n_e$ is
obviously lower at a shorter  $\lambda_L$. This is the reason why
LPI levels are lower at 3$\omega$ and 4$\omega$. As it is shown,
$n_e$ of model IV is  close to that of model III at $n_e \ge$  0.02,
again because they both use $2\omega$ laser during the main-pulse.
Notice that a lower LPI backscatters of model IV can be expected
than  model III, because  they have a  similar  spacial distribution
of $n_e$ while model IV has a higher $T_e$ than model III.
Nevertheless, as we have mentioned above, the LPI levels needs to be
measured by experiments for the quantitatively study.

The reason for  the decrease of $E_r$ inside the foam-solid Au has
not been discussed in previous publications, but it is not
surprising. Presented in Fig.\ref{optical depth} is the temporal
evolutions of the optical depth for all models. As shown, optical
depth of the foam-solid is remarkably decreased about 4 to 7 times
as compared to the solid models, and this is due to its shorter
expansion and higher $T_e$ than the solid models.

\begin{figure}[htbp]
\includegraphics[bb=170 150 420 550,width=4 cm]{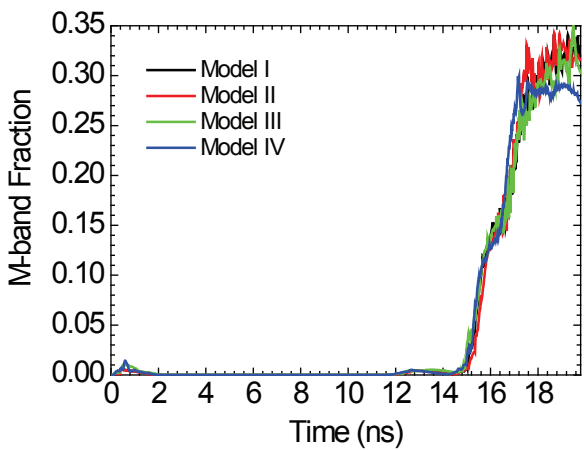}
\caption{(color online) M-band fractions of the output  x-ray flux
for all models.
} \label{Mband}
\end{figure}

Finally, it is worth to compare the M-band fraction $f_M$ between
the foam-solid target and the solid target, because the x-ray
spectrum is very important for the implosion
performances\cite{Olson}. The portion of the X-ray flux above  1.8
keV generated in the laser-heated Au can preheat capsule and arouse
hydrodynamic instabilities \cite{lindl2003, Huo2012}.  In our
simulations, we define the emissions between 1.8 keV and 4 keV as
the M-band of Au. Presented in Fig. \ref{Mband} is temporal
evolutions of $f_M$ for all models. As shown, $f_M$ is very small
before 14 ns, but later it increases rapidly with  laser power and
reaches about $15\%$ at around 16 ns for all models. For the first
three models with the solid Au, $f_M$ varies as $\lambda_L$ while
the differences are small. However, $f_M$ is more sensitive to
$\rho_0$. For model IV which uses the foam-solid target, $f_M$
obviously decreases as compared to the results of the solid Au. At
around 19 ns, $f_M$ is around $34\%$ in the first three models,
while it is around $29\%$ in models IV. Therefore, $f_M$  can be
clearly decreased with the foam-solid target.

In this work, we also simulate a model in which the foam-solid Au is
driven by a 2$\omega$ laser during whole ignition pulse and compare
its simulation results with  model IV. The comparisons show that
their results are close, which is reasonable because they all use
2$\omega$ for the main-pulse. In addition, we also consider the
foam-solid targets with  $\rho_0$ = 0.25 g/cm$^3$ and $\Delta$ = 80
$\mu$m or  $\rho_0$ = 1 g/cm$^3$ and $\Delta$ = 20 $\mu$m for the
foam part, and their simulation results have no big differences from
what we have for model IV. In fact, to search the optimum $\rho_0$
of foam for ignition hohlraum design, one  should study and compare
the x-ray radiations inside the hohlraums with various foam density
under an ignition laser drive by using 2D or 3D simulations and
performing systematic experiments with both modest energy (SGIII and
Omega) and MJ scale lasers such as the NIF and LMJ. In future
studies,  a thick foam should be used, which thickness can cover
both laser ablated region and radiation ablated region.

\section{SUMMARY}

We have proposed to drive a foam  hohlraum wall  with an ignition
laser pulse by the 4$\omega$-2$\omega$  ignition  pulse,
 in order to take the advantage of  4$\omega$
laser in its low LPI   during the pre-pulse, the advantage of
2$\omega$ laser in its relatively high energy output and low damage
threshold during the main-pulse, and the advantage of   foam in its
high  $\eta_{LX}$, high  $\eta_{HC}$ and relatively low M-band
fraction in its radiation spectrum. The advantage of  foam in a
higher   $\eta_{HC}$ was predicted and demonstrated in previous
publications. In this work, we  have demonstrated the advantage  of
  foam in a higher  $\eta_{LX}$ by comparing the laser ablations
between a solid target and a foam-solid target under an ignition
laser pulse with our 1D simulations. As a result, $\eta_{LX}$ can be
increased by $28\%$ in the foam-solid Au under the
4$\omega$-2$\omega$ pulse as compared to a pure solid Au  under a
3$\omega$ laser. Our studies have shown that the increase of
$\eta_{LX}$ of the foam-solid Au is due to the decreases of its
kinetic energy and radiation energy as compared to a solid Au, and
the decrease of the radiation energy is caused by the shorter
optical depth inside a foam. In addition, a lower LPI levels is
expected from  foam  than from   solid  because of its higher
electronic temperature in the laser deposited region.

Because 2$\omega$ laser energy delivered to drive a fusion target
can exceed  3$\omega$ by a factor of about 1.1 to 1.2, together with
the increased $\eta_{LX}$, $\eta_{HC}$ and the decreased LPI levels
inside the foams, it may provide a  larger and flexible design space
for ignition with a foam hohlraum wall and a 4$\omega$-2$\omega$
laser pulse. Because of the low damage threshold of the 2$\omega$
laser at high power laser, we can therefore expect a remarkable
increase of shot number at high laser energy power on a laser
ignition facility for ignition study. Nevertheless, it is mandatory
to have experiments for an understanding and demonstration of
acceptable LPI levels and target coupling physics of  2$\omega$
lasers.


\begin{acknowledgements}
The authors wish to acknowledge the beneficial help from Guoli Ren,
Yaohua Chen, Yongsheng Li, and Wanguo Zheng. This work is supported
by the Development Foundation of CAEP (Grant No. 2013A0102002) and
the National Natural Science Foundation of China (Grant No.
11475033).

\end{acknowledgements}


\begin{thebibliography}{22}


\bibitem{lindl2003} J. D. Lindl, B. A. Hammel, and B. G. Logan, D. D. Meyerhofer,
S. A. Payne and J. D. Sethian, Plasma Phys. Control. Fusion {\bf 45}, A217-A234 (2003).




\bibitem{lindl2014} J. D. Lindl, O. Landen, J. Edwards, E. Moses, and NIC Team,
``Review of the National Ignition Campaign 2009-2012", Phys. Plasmas
{\bf 21}, 020501 (2014).

\bibitem{zhengwanguo}  W. Zheng, X. Wei, Q. Zhu, F. Jing, D. Hu, J. Su, K. Zheng, X. Yuan, H. Zhou,
W. Dai, W. Zhou, F. Wang, D. Xu, X. Xie, B. Feng, Z. Peng, L. Guo,
Y. Chen, X. Zhang, L. Liu, D. Lin, Z. Dang, Y. Xiang, and X. Deng,
High Power Laser Science and Engineering {\bf 4},
doi:10.1017/hpl.2016.20 (2016).

\bibitem{lanke2016}  K. Lan, J. Liu, Z. Li, X. Xie, W. Huo, Y. Chen,
G. Ren, C. Zheng, D. Yang, S. Li, Z. Yang, L. Guo, S. Li, M. Zhang,
X. Han, C. Zhai, L. Hou, Y. Li, K. Deng, Z. Yuan, X. Zhan, F. Wang,
G. Yuan, H. Zhang, B. Jiang, L. Huang,  W. Zhang, K. Du, R. Zhao, P.
Li, W. Wang, J. Su, X. Deng, D. Hu, W. Z, H. Jia, Y. Ding, W. Zheng,
X. He, Matter and Radiation at Extremes {\bf 1}, 8 (2016).

\bibitem{Campbell} E. M. Campbell, R. Cauble, and B. A. Remington, AIP Conference Proceedings {\bf 429}, 3 (1998).



\bibitem{LMJ} J. Giorla, J. Bastian, C. Bayer, B. Canaud, M. Casanova,
F. Chaland, C. Cherfils, C. Clique, E. Dattolo, P. Fremerye, D.
Galmiche, F. Garaude, P. Gauthier, S. Laffite, N. Lecler, S.
Liberatore, P. Loiseau, G. Malinie, L. Masse, A. Masson, M. C.
Monteil, F. Poggi, R. Quach, F. Renaud, Y. Saillard, P. Seytor, M.
Vandenboomgaerde, J. Van der Vliet, and F. Wagon,Plasma Phys.
Control. Fusion {\bf 48} B75 (2006).

\bibitem{Neumayer} P. Neumayer, R. L. Berger, D. Callahan, L. Divol,
 D. H. Froula, R. A. London, B. J. MacGowan, N. B. Meezan, P. A. Michel,
 J. S. Ross, C. Sorce, K. Widmann, L. J. Suter, and S. H. Glenzer, Phys. Plasmas {\bf 15}, 056307 (2008).

\bibitem{Kline2011} J. L. Kline, S. H. Glenzer, R. E. Olson, L. J. Suter,
K. Widmann, D. A. Callahan, S. N. Dixit, C. A. Thomas, D. E. Hinkel,
E. A. Williams, A. S. Moore, J. Celeste, E. Dewald, W. W. Hsing, A. Warrick,
J. Atherton, S. Azevedo, R. Beeler, R. Berger, A. Conder, L. Divol,
C. A. Haynam, D. H. Kalantar, R. Kauffman, G. A. Kyrala, J. Kilkenny,
J. Liebman, S. Le Pape, D. Larson, N. B. Meezan, P. Michel, J. Moody,
M. D. Rosen, M. B. Schneider, B. Van Wonterghem, R. J. Wallace, B. K. Young,
O. L. Landen, and B. J. MacGowan, Phys. Rev. Lett. {\bf 106}, 085003 (2011).

\bibitem{Froula2007} D. H. Froula, L. Divol, N. B. Meezan, S. Dixit, J. D. Moody,
P. Neumayer, B. B. Pollock, J. S. Ross, and S. H. Glenzer, Phys. Rev. Lett. {\bf 98}, 085001 (2007).

\bibitem{Haynam} C. A. Haynam, P. J. Wegner, J. M. Auerbach, M. W. Bowers,
S. N. Dixit, G. V. Erbert, G. M. Heestand, M. A. Henesian, M. R. Hermann,
K. S. Jancaitis, K. R. Manes, C. D. Marshall, N. C. Mehta, J. Menapace,
E. Moses, J. R. Murray, M. C. Nostrand, C. D. Orth, R. Patterson,
R. A. Sacks, M. J. Shaw, M. Spaeth, S. B. Sutton, W. H. Williams, C. C. Widmayer,
R. K. White, S. T. Yang, and B. M. Van Wonterghem, Appl. Opt. {\bf 46}, 3276 (2007).

\bibitem{Suter} L. J. Suter, S. Glenzer, S. Haan, B. Hammel, K. Manes, N. Meezan, J. Moody, M. Spaeth, and L. Divol K. Oades and M. Stevenson, Phys. Plasmas {\bf 11}, 2738 (2004).

\bibitem{Niemann} C. Niemann, R. L. Berger, L. Divol, D. H. Froula, O. Jones, R. K. Kirkwood, N. Meezan, J. D. Moody, J. Ross, C. Sorce, L. J. Suter, and S. H. Glenzer, Phys. Rev. Lett. {\bf 100}, 045002 (2008).

\bibitem{Depierreux} S. Depierreux, D. T. Michel, V. Tassin, P. Loiseau, C. Stenz, and C. Labaune, Phys. Rev. Lett. {\bf 103}, 115001 (2009).

\bibitem{Stevenson} R. M. Stevenson, K. Oades, B. R. Thomas, M. Schneider, G. E. Slark, L. J. Suter, R. Kauffman, D. Hinkel, and M. C. Miller,  Phys. Rev. Lett. {\bf 94}, 055006 (2005).

\bibitem{Labaune} C. Labaune,  Nature Phys.  {\bf 3}, 680 (2007).



\bibitem{Depierreux2012} S. Depierreux, P. Loiseau, D. T. Michel, V. Tassin, C. Stenz, P.-E. Masson-Laborde, C. Goyon, V. Yahia, and C. Labaune, Phys. Plasmas {\bf 19}, 012705 (2012).

\bibitem{Lizhichao} Z. Li, J. Zheng, X. Jiang, Z. Wang, D. Yang, H. Zhang,
S. Li, Q. Yin, F. Zhu, P. Shao, X. Peng, F. Wang, L. Guo, P. Yuan,
Z. Yuan, L. Chen, S. Liu, S. Jiang, and Y. Ding, Phys. Plasmas {\bf
19}, 062703 (2012).



\bibitem{HR2005} M. D. Rosen and J. H. Hammer, Phys. Rev. E. {\bf 72}, 056403 (2005).

\bibitem{HR2008} P. E. Young, M. D. Rosen, J. H. Hammer, W. S. Hsing, S. G. Glendinning, R. E. Turner, R. Kirkwood, J. Schein, C. Sorce, J. H. Satcher, Jr., A. Hamza, R. A. Reibold, R. Hibbard, O. Landen, A. Reighard, S. McAlpin, M. Stevenson, and B. Thomas, Phys. Rev.
Lett. {\bf 101}, 035001 (2008).

\bibitem{Dongyunsong2013} Y. Dong, W. Shang, J. Yang, L. Zhang, W. Zhang, Z. Li, L. Guo, X. Zhan, H. Du, B. Deng, and Y. Pu,  Phys. Plasmas {\bf 20}, 123305 (2013).

\bibitem{Zhanglu2015} L. Zhang, Y. Ding, S. Jiang, J. Yang, H. Li, L. Kuang, Z. Lin, L. Jing, L. Li, B. Deng, Z. Yuan, T. Chen, G. Yuan, X. Tan, and P.Li,  Phys. Plasmas {\bf 22}, 110703 (2015).

\bibitem{Zhanglu2016} L. Zhang, Y. Ding, Z. Lin, H. Li, L. Jing, Z. Yuan, Z. Yang, X. Tan, L. Kuang, W. Zhang, L. Li, P. Li, G. Yuan, S. Jiang and B. Zhang,  Nucl. Fusion {\bf 56}, 036006 (2016).

\bibitem{Zhanglu2011} L. Zhang, Y. Ding, J. Yang, S. Wu, and S. Jiang, Phys. Plasmas {\bf 18}, 033301 (2011).

\bibitem{Dongyunsong2013-1}  Y. Dong, L. Zhang, J. Yang, and W. Shang,  Phys. Plasmas {\bf 20}, 123102 (2013).




\bibitem{Feng1999} T. Feng, D. Lai,  and Y. Xu,  Chinese J. Comput. Phys. {\bf 16}, 199-205 (1999).

\bibitem{Xuyan2006}  Y. Xu, S. Jiang, D. Lai, W. Pei, Y. Ding, T.
Chang, K. Lan, S. Li, and T, Feng,  Laser and Particle Beams {\bf
24}, 495(2006)

\bibitem{Huo2012} Y. Li, W. Y. Huo and K. Lan,  Phys. Plasmas {\bf 18}, 022701 (2011);
W. Y. Huo, K. Lan, Y.  Li, D. Yang, S. Li, X. Li, C. Wu, G. Ren, Y.
Zhao, S. Zou, W. Zheng, P. Gu, M. Wang, R. Yi, X. Jiang,  T. Song,
Z. Li, L. Guo, Y. Liu, X. Zhan, F. Wang, X. Peng, H. Zhang,  J.
Yang, S. Liu, S. Jiang, and Y. Ding, Phys. Rev. Lett. {\bf 109},
145004(2012).

\bibitem{Xuyan} Y. Xu, S. Wu,  and W. Zheng,
Phys. Plasmas {\bf 22}, 042708 (2015).



\bibitem{Feng1998} T. Feng, D. Lai,  Y. Xu, and J. Li, {\it Annual Reports of China Academy of Engineering Physics
1998} (Atomic Energy Press, Chengdu, 1998), p. 286 (in Chinese).


\bibitem{Cohen} R. S. Cohen, L. Spitzer, and P. M. Routly, Phys. Rev.\textbf{ 80}, 230
(1950).

\bibitem{Spitzer} L. Spitzer and R. H$\ddot{a}$rm, Phys. Rev.\textbf{ 89}, 977 (1953).

\bibitem{Malone}R. C. Malone, R. L. McCrory, and R. L. Morse, Phys. Rev. Lett. \textbf{12},
721 (1975).

\bibitem{Haan}S. W. Haan, J. D. Lindl, D. A. Callanhan, D. S. Clark, J. D. Salmonson,
B. A. Hammel, L. J. Atherton, R. C. Cook, M. J. Edwards, S. Glenzer,
A. V. Hamza, S. P. Hatchett, M. C. Herrmann, D. E. Hinkel, D. D. Ho,
H. Huang, O. S. Jones, J. Kline, G. Kyrala, O. L. Landen, B. J.
MacGowan, M. M. Marinak, D. D. Meyerhofer, J. L. Milovich, K. A.
Moreno, E. I. Moses, D. H. Munro, A. Nikroo, R. E. Olson, K.
Peterson, S. M. Pollaine, J. E. Ralph, H. F. Robey, B. K. Spears, P.
T. Springer, L. J. Suter, C. A. Thomas, R. P. Town, R. Vesey, S. V.
Weber, H. L. Wilkens, and D. C. Wilson,  Phys. Plasmas {\bf 18},
051001 (2011).


\bibitem{MTV} S. Atzeni, J. Meyer-ter-Vehn, The Physics of Inertial Fusion (Oxford Science, Oxford, 2004).

\bibitem{Kline056314} J. L. Kline, D. A. Callahan, S. H. Glenzer, N. B. Meezan, J. D. Moody, D. E. Hinkel, O. S. Jones, A. J. MacKinnon, R.
Bennedetti, R. L. Berger, D. Bradley, E. L. Dewald, I. Bass, C.
Bennett, M. Bowers, G. Brunton, J. Bude, S. Burkhart, A. Condor, J.
M. Di Nicola, P. Di Nicola, S. N. Dixit, T. D¡§oeppner, E. G.
Dzenitis, G. Erber, J. Folta, G. Grim, S. Lenn, A. Hamza, S. W.
Hann, J. Heebner, M. Henesian, M. Hermann, D. G. Hicks, W. W. Hsing,
N. Izumi, K. Jancaitis, O. S. Jones, D. Kalantar, S. F. Khan, R.
Kirkwook, G. A. Kyrala, K. LaFortune, O. L. Landen, L. Lain, D.
Larson, S. Le Pape, T. Ma, A. G. MacPhee, P. A. Michel, P. Miller,
M. Montincelli, A. S. Moore, A. Nikroo, M. Nostrand, R. E. Olson, A.
Pak, H. S. Park, M. B. Schneider,
M. Shaw, V. A. Smalyuk, D. J. Strozzi, T. Suratwala, L. J. Suter, R. Tommasini, R. P. J. Town, B. Van Wonterghem, P. Wegner, K. Widmann, C. Widmayer, H. 
Phys. Plasmas {\bf 20}, 056314 (2013).

\bibitem{Olson}R. E. Olson, R. J. Leeper, A. Nobile, and J. A. Oerte, Phys. Rev.Lett. {\bf 91}, 235002 (2003).








\end{thebibliography}
\end{document}